\documentstyle[preprint,revtex,eqsecnum]{aps}
\begin{document}
\tolerance 10000
\draft

\vskip 6 truecm
September, 1993 \hskip 7 truecm  Report LPQTH-93/18

\begin{center}
{\bf PERSISTENT CURRENTS IN 1D DISORDERED RINGS OF INTERACTING
ELECTRONS}
\end{center}

\vspace*{0.4cm}
\author{G.BOUZERAR\cite{byline1}, D. POILBLANC\cite{byline2}}
\author{and}
\author{G. MONTAMBAUX\cite{byline3}}
\vspace*{0.3cm}
\begin{instit}
\begin{center}
Groupe de Physique Th\'eorique,\\
Laboratoire de Physique Quantique,\\
Universit\'{e} Paul Sabatier,\\
31062 Toulouse, France
\end{center}
\end{instit}
\vspace*{0.2cm}

\receipt{\hskip 4truecm}

\begin{abstract}
We calculate the persistent current of 1D rings of spinless fermions
with short-range interactions on a lattice with up to 20 sites, and in the
presence of disorder, for various band fillings.
We find that {\it both} disorder and interactions always decrease the
persistent current by localizing the electrons.
Away from half-filling, the interaction
has a much stronger influence in the presence of disorder than in the pure
case.
\end{abstract}
\vspace*{2cm}

\pacs{PACS numbers: 72.10.-d, 71.27.+a, 72.15.Rn }

The recent discovery of persistent currents in mesoscopic rings has
addressed new interesting questions on the thermodynamics of these systems.
Although such an effect was predicted for a long time, the
unexpectedly large amplitude of the measured currents lead to important
interrogations.
Among them, the role of e-e interactions is still unclear. It has been
proposed \cite{ambegaokar} that the interactions contribute to the average
current which is measured in a many rings experiments \cite{bell}.
On the other hand, the importance of the choice of the statistical ensemble
to calculate average values has also been stressed  \cite{average}.
Although the first explanation, based on a perturbative calculation both in
interaction and disorder, seems to give a quantitative estimate
closer
to the experiment, it  is still  too small by one order of magnitude
and the interaction parameter used in the theory  is not well known.

In addition, for a single ring experiment \cite{ibm}, the magnitude of the
measured current is also not understood
and, up to now, perturbation theory has failed to explain an enhancement of
the current \cite{perturbation}.  It is only
when disorder is weaker that experiment and theory seem to
agree \cite{benoit}, even for non interacting electrons.

At the moment, the role of the interactions in disordered systems is still
unclear and the subject is vastly open.
It has been recently proposed that, in the presence of
interactions, the current should be larger than for
free electrons, the effect of the interactions
being to counteract the disorder effect \cite{weidenmuller}.

The aim of this paper is to describe the interplay between the
interactions {\it and}
the disorder on the persistent currents in 1D rings.
We choose a model of spinless fermions with short range interactions on a
lattice. Our main result is that a repulsive  interaction always decreases
the amplitude of the current.
It is well known that the 1D description is certainly not the most appropriate
one to  describe quantitatively experiments which are performed in rings
with finite width, in the diffusive regime.
But our hope is to find numerical results which may give indications for a
more real situation.

We describe a chain of  1D spinless
fermions in the presence of disorder with the following Hamiltonian :
\begin{eqnarray}
{\cal H} =
-t/2 \sum_{\bf i}
(\exp{(2i\pi\Phi/N)}\,c^{\dagger}_{\bf i}\,
 c_{\bf i+1}
+ h.c.) +V\sum_{\bf i}n_{\bf i}\,n_{\bf i+1} +
\sum_{\bf i}w_{\bf i}\,n_{\bf i}
\label{tVW}
\end{eqnarray}
\noindent
where $w_{\bf i}$ are on-site energies and are chosen randomly between -W/2
and W/2 and V is the nearest neighbour Coulomb repulsion. In the following we
will
 take t=1, $\Phi$ is the total  magnetic flux through the ring (measured in
units of
flux quantum $\Phi_{0}=h/e $) and N is the number of sites.
(We use the conventional notation for the amplitude of disorder W and for
the nearest neighbors interaction V. These notations are opposite to those
used in ref.\cite{weidenmuller}).

Let us first recall some  physical properties of this hamiltonian without
disorder, i.e. W=0.
In one dimension, for repulsive
interaction, a metal-insulator transistion occurs at half-filling due to the
existence of umklapp processes.
However away from half-filling, the umklapp processes become irrelevant and the
system
is expected to be metallic
\cite{shankar}.
For Hamiltonian (\ref{tVW}), the metal-insulator (Mott) transition
occurs at V=1.
The system is insulator for $V > 1$ and metallic for $ 0 \le V < 1 $ .

This transition can also be described in the spin picture\cite{shankar}.
In the case $W=0$, the model is integrable and is formally
equivalent to an anisotropic spin model, as obtained by a
standard Wigner-Jordan transformation.
By this way the new Hamiltonian reads:
\begin{eqnarray}
{\cal H}_{XXZ} =
-t/2\sum_{\bf i}
(\exp(2i\pi\Phi/N)\,S^{+}_{\bf i}\,
 S_{\bf i+1}^{-}
+ h.c.) +V[N/4 +\sum_{\bf i}S_{\bf i}^{z}\,S_{\bf i+1}^{z}]
\label {tV}
\end{eqnarray}
\noindent
For the XXZ model, V=1 corresponds to a transition from an XY-model ($ V <1 $)
to an Ising model ($V >1$). $V=1$ corresponds to an isotropic Heisenberg
system.
Note that a spin gap opens up for $V >1$ .
It corresponds to the gap in the charge excitations for Hamiltonian (\ref{tVW})
characteristic of the insulator.
With disorder, $W\neq0$, we must add in (\ref{tV}), the following term
\begin{eqnarray}
{\cal H}_{random}= \sum_{\bf i}w_{\bf i}\,(S_{\bf i}^{z} -1/2)
\end{eqnarray}
\noindent
This term describes the interaction of
the local spins with random magnetic fields.

We now turn to the numerical calculation of the ground state
energy as a function of the total magnetic flux $E(\Phi)$ which is
the first step of our work.
The ground state energy is obtained from  a standard Lanczos algorithm
\cite{lanczos}.
First of all, let us briefly describe some technical aspects of the method.
The calculation is limited to relatively small system sizes N, since
(i) the Hilbert space dimension grows exponentially  fast with N,
(ii)  we have to average over many realizations of the disorder, (because of
statistical fluctuations) and
(iii) the disorder breaks the translation symmetry.
The system sizes we were interested in, vary from 6 to 20 sites
and we have chosen to consider two different cases: half
and quarter fillings.
Let us remind that the Lanczos method consists in the construction of a
tridiagonal matrix by applying iteratively
the Hamiltonian on an initial random vector. By this way a
basis of normalized vectors $\Psi_{n}$ is defined as well as
a set of values $e_{n}$ and $b_{n}$ given by the relationship,
$ {\cal H}\,\Psi_{n}=b_{n-1}\Psi_{n-1}+e_{n}\Psi_{n}+b_{n+1}\Psi_{n+1} $.
Hence we construct by iteration a tridiagonal Hamiltonian matrix expressed in
the $\Psi_{n}$ basis
that we diagonalize to obtain the spectrum of the eigenvalues.
This kind of process is rapidly converging.

Let us, now, consider the calculation of the persistent current in such
rings threaded by a total flux $\Phi$.
As usual the current is defined by
\begin{eqnarray}
            I(\Phi)=-\frac{1}{2\pi}\frac{\partial E(\Phi)}{\partial\Phi},
\end{eqnarray}
\noindent
where $E(\Phi)$ is calculated by exact diagonalization of the
Hamiltonian.
As well known, the flux can be gauged out from the Hamiltonian so that the
presence of an Aharonov-Bohm flux through a ring is analogous to a twist in
the boundary conditions $ \Psi(x+N)=\Psi(x)e^{2i\pi\Phi }$. The spectrum and
the persistent current have the flux periodicity of one.

In figs.1,2,  $I(\Phi)$ is plotted versus $\Phi$ for a 16 site
ring at electron density $<n>=0.5$ and $<n>=0.25$ in (a) and
(b) respectively, for different values
of the disorder and the interaction.
Fig.1 corresponds to the 'ordered' interacting case $W=0$ and
disorder $W=0.5$ is introduced in fig.2.
On fig.1, a discontinuity of $I(\Phi)$ appears
at $\Phi=0$, for zero disorder ($W=0$).
Indeed, in the absence of disorder, translation symmetry
is preserved and total momentum is a good quantum number.
As a function of $\Phi$, a crossing occurs between two lowest energy levels
with
different momenta.
This crossing occurs at $\Phi=0$ for an even number of electrons
or at $\Phi=0.5$ for an odd number.
This can be easily understood in the non interacting case where
$E(\Phi)=-$ $\sum_{\bigl(n\in {\cal E}(n)\bigr)}$ $cos(k_{n}+2\pi\Phi/N)$ and
the subset of
the electron momenta $k_{n}=2\pi\,n/N$ is chosen in order to minimize the
total energy. Since translation invariance is preserved in the
presence of interactions, the discontinuity still exists for finite $V$.
When disorder is introduced ($W\neq 0$) in fig.2, the scattering potential
lifts the
degeneracy at the
crossing point and hence leads to a continuous variation of
the current.
In fig.1a (half-filling) we clearly observe the effect of the Mott transition
on the currents:
we notice that for $V<1$, I is slowly varying with V, but when $V>1$
a drop of the current appears.
However, away from half-filling (see fig.1b), the current is not affected
for moderate  interactions.
At $W=0$ and away from half-filling, the system is always metallic.

In figs.2a and b, the influence of the repulsive interaction is shown for a
fixed impurity
potential of magnitude $W=0.5$ and for the same parameters as in figs.1
(N=16, $<n>=0.5$ and $<n>=0.25$).
Clearly, in the half-filled case (fig.2a), the repulsion tends to suppress the
current even further.
This is reminicent of the Mott localization which occurs in the pure system.
With increasing W, $I(\Phi)$ decreases as expected due to a stronger
localization by the impurity potential.
Such an effect has also been found in a ring of spinless fermions with
long-range interactions\cite{abraham}.

More interesting is the effect of
the interaction {\it away} from half-filling where no localization is expected
in the absence of disorder.
As seen previously in fig.1b the effect of the interaction for $W=0$ is
extremely
weak, because of the absence of Umklapp
scattering.
However, it is clear from fig.2b that the interaction is much more effective in
the presence
of the disorder ($W\neq 0$); while on fig.1b V had almost no effect, in fig.2b,
for a relatively weak disorder, V leads to a significant {\it decrease} of the
current.
Such a striking influence of the interaction V is also seen in the metallic
regime at half-filling (compare e.g. $V=0$ and 1 in figs.1a and 2a).

At this point, we would like to describe more qualitatively
the transition from the localized regime (insulator) to the ballistic one
(perfect
metal).
As stressed by Scalapino and al. \cite{dru}, the Drude weight $ \pi D$ is a
relevant parameter to characterize both of them.
As originally noted by Kohn \cite{Kohn}, the Drude weight $ \pi D$ can be
calculated from the dependence of the ground state energy versus $\Phi$ ,

\begin{eqnarray}
    D={N \over 4 \pi^2}(\frac{\partial^2
E(\Phi)}{\partial\Phi^2})_{\Phi=\Phi_{m}}. \label{pdru}
\end{eqnarray}
\noindent
where $\Phi_{m}=0$ or 1/2 is the location of the minimum of $E(\Phi)$.
As mentionned earlier, $\Phi_{m}$
depends on the parity of the number of electrons.
For an even electron number parity, we take D as the second
derivative at $\Phi_{m}$= 1/2.

Note  that for free electrons
$D = <n> / m$
where $<n>$ is the density of the mobile charge carriers and $m = 1 / 2t$ is
their mass. Generally D is given by \cite{scalapino} :

\begin{eqnarray}
 D = \frac{<n>}{m^*}
\end{eqnarray}
\noindent
where $m^*$ is the effective mass of the carriers renormalized by the
interaction.
By this way, the D parameter can determine the different regimes.
A perfect metal (ballistic regime) will be caracterized by a finite value of
D. This
corresponds to a persistent current $I$ scaling as $1/N$. In the insulator,
D vanishes exponentially
as the size of the system goes to infinity, $D\propto\,e^{-N/\xi}$,
where $\xi$ is the localization length.
As a check of the numerical calculations on the drude peak, we observed the
correct finite size scaling for $W=0$, at half-filling
$D \sim D_{lim} + a/N^{2} $ for $V<1$ and
$D\propto\,e^{-N/\xi}$ for $V>1$ when $N > \xi$.

Since we consider disordered systems, we have to average
over many realizations of the disorder.
The number of configurations we averaged over vary from
50 to 250, depending on the size of the Hilbert space
and the filling.
In fig.3 ($<n>=0.5$) and fig.4 ($<n>=0.25$) D is plotted versus 1/N (N is
the size of the ring).
We first consider the half-filled case (fig.3). When V=0 and W=0,
D goes to a finite value in the thermodynamic
limit ($D_{lim}=1/ \pi$).
As long as W=0 (i.e. without disorder),
the Drude weight is weakly affected by a small
interaction  $ V<1 $, signature that the system remains metallic.
But for $ V>1 $ (here V=2) D decreases faster with N, the system becomes an
insulator. We now turn to the effect of disorder.
Once $W \neq 0$,  we observe a tendancy towards
localization for all V. With disorder in the system, the effect of
the interaction is also to increase the degree of localization.
In the light of the results given by fig.3 and 4, it appears
that the role played by the interaction in presence of disorder ($W\neq0$) is
clearly
different for $<n>=0.5$ and for $<n>=0.25$.
On one hand, at $<n>=0.5$ we see that the interaction V
leads to a real decrease.
For example, if we consider in fig.3 the case $W=2$ and compare the data for
$V=0$
and 2 we notice that the presence of the interaction reduces
the localization length $\xi $ by more than a factor 3.
On the other hand, at $<n>=0.25$ (and fixed W) the value of D is
less affected by the interaction V, as also observed on figs.2a and 2b for the
current.
However, we note that the effect of V is significantly larger for $W\neq 0$
than
in the pure case $W=0$.
In summary, we do not observe any increase of D due to
the competition between the interaction and the disorder.

We finish this paper by a few remarks on the conductivity spectrum.
The optical conductivity is given by ,

\begin{eqnarray}
    \sigma(\omega)= \pi D \delta(\omega)+\sigma_{reg}(\omega)
\end{eqnarray}
\noindent
where $\sigma_{reg}(\omega)$ given by the Kubo formula

\begin{eqnarray}
\sigma_{reg}(\omega)=\frac{\pi}{N}\sum_{m\neq0}\frac{|<m|\hat{j}|0>|^2}{E_{m}-E_{0}}\delta(\omega-(E_{m}-E_{0}))
\end{eqnarray}
\noindent
and $\hat{j}$ is the current density operator
\begin{eqnarray}
\hat{j}=-it/2 \sum_{\bf i}(
{c^{\dagger}_{\bf i}\,
 c_{\bf i+1}}\,exp(2i\pi\Phi/N)- hc)
\label{cour}
\end{eqnarray}
\noindent
All quantities in (9) are calculated at $\Phi=\Phi_{m}$ and $E_{n}$ are the
excited manybody energies.
The amplitude D of the Drude $\delta(\omega)$ peak was calculated previously.
We have explicitely checked the sum rule,
\begin{eqnarray}
\int_{0}^{\infty}\sigma(\omega)d\,\omega=-(\pi/2N)<0|H_{kin}|0>
\end{eqnarray}
\noindent
where $<0|H_{kin}|0>$ is the groundstate expectation value of the kinetic
energy.

In fig.5, $\sigma(\omega)$ is plotted versus $\omega$ for a 16 site ring
at half-filling, with $V=0.5$ and for $W=0$ or $W=1$.
In presence of the disorder we had averaged over 100 configurations.
We clearly see, as expected, that in the metallic case $W=0$, the contribution
at non zero frequency
is negligeable ($\pi D \sim 0.98 $ and only less than 1\%  of the weight
is left at finite frequencies).
However, when we introduce disorder, a strong absorption appears
at non zero frequency with a peak around $\omega=0.4t$, while weight
is removed from $\omega=0$ ($\pi D \sim 0.52$).

This broad absorption can be interpreted physically in the following way:
if we assume that disorder localizes the
wavefunctions in small finite size regions
with a broad distribution of volumes, this leads to a correspondingly broad
distribution of finite characteristic frequencies.
In other words, the localized electrons can oscillate in disconnected regions
of
different sizes.

Let us summarize the main result of this paper. In our model for interacting
electrons, we never observe an increase
of the persistent current when interaction is switched on.
At half-filling, the interaction induces a metal-insulator transition . The
current strongly decreases, in qualitative agreement with the result of
ref.\cite{abraham}.
Away from half-filling, the effect of the interaction is much weaker in the
absence of disorder. However when impurity scattering exists, the interaction
plays again a crucial role and leads to
an additional decrease of  the current. This is because it is more
difficult to move correlated electrons in a random potential than
independent electrons.
Our results are in
discrepancy with those
of \cite{weidenmuller}. It will be of interest to know if the qualitative
results obtained in this paper still apply for a multichannel ring.

Support from the Centre de Calcul Vectoriel
pour la Recherche (CCVR), Palaiseau, France is acknowledged. Laboratoire
de Physique Quantique (Toulouse) and Laboratoire de Physique des Solides
(Orsay) are Unit\'es Associ\'ees No. URA505 and URA2 du CNRS respectively.

\newpage
\bigskip
\centerline{FIGURE CAPTIONS}
\medskip

\noindent
{\bf Figure 1}

\noindent
Current $I(\Phi)$ versus $\Phi$ for a 16 site ring at $<n>=0.5$ (fig.1a) or
at $<n>=0.25$ (fig.1b), fixed $V=0$ and
$W=0$, 0.5 and 1.

\noindent
{\bf Figure 2}

\noindent
Current $I(\Phi)$ versus $\Phi$ for a 16 site ring at $<n>=0.5$ (fig.2a) or
$<n>=0.25$ (fig.2b)
,at fixed $W=0$
and $V=0$, 1 and 2.

\noindent
{\bf Figure 3}

\noindent
Scaling of D at half-filling ($<n>=0.5$). D vs 1/N for different
V and W.

\noindent
{\bf Figure 4}

\noindent
Scaling of  D at quarter filling ($<n>=0.25$).

\noindent
{\bf Figure 5}

\noindent
Total conductivity $\sigma(\omega)$ for a 16 site ring at $<n>=0.5$ and (a)
with
or (b)
without disorder ($W=0$ and $W=1$) for a fixed $V=0.5$.
For $W=0$ $\pi D \sim 0.98$.
For $W=1$, $\pi D_{aver.} \sim .52$ and we averaged over 100 realisations
of the disorder.
\end{document}